# VERTICAL MAGNETIC FIELD ON THE OUTER BOUNDARY OF SUNSPOT UMBRA

## Efremov V.I., Parfinenko L.D., Soloviev A.A.

*Central (Pulkovo) astronomical observatory of RAS, St. Petersburg, Russia*

**1. Introduction**
Recently, several papers have appeared [1–3], in which various regimes of magnetoconvection in the shadow and penumbra of a sunspot are discussed. The central issue in them was the drawing of the "shadow-penumbra" (U-PU) boundary and the determination of the magnitude of the vertical magnetic field on it. For example, it was found in [1] that the vertical magnetic component of the field Bver changes little along the U-PU boundaries, and the average value of this field at the U-PU boundary is 1867 G (Hinode / SP). Moreover, this value, according to [1], does not depend on either the size (area) of the sunspot or the cycle of solar activity, in fact, it is some universal physical constant for all sunspots. We intend to test this far-reaching conclusion using an independent mathematical method.

**2. Data and analysis**
New Solar Dynamics Observatory (SDO) Helioseismic and Magnetic Imager (HMI) instrument data are now available, referred to as Space-weather HMI Active Region Patches (SHARPs) (jsoc.stanford.edu, [4]) with a 12-minute cadence. SHARP data are available in two coordinate systems (jsoc.stanford.edu, [5]; [6]): CCD (31 segments) and CEA (11 segments, where all values are remaped in heliographic In this work, we use the CEA coordinate system and 2 data segments: continuum (Ic) and magnetic field (Br). Transformation used in SHARP data Lambert "straightens out" the projection effects by bringing the object (Patch) to the central meridian, but we nevertheless, in order to avoid possible distortions, chose the spots at the moment they passed the central meridian. So, to determine the values of the vertical component of the magnetic field (Bver) at the U-PU boundary, one must first determine the coordinates of the contour of this boundary. Then this contour is transferred to the magnetic field map, the field change data is read from it, the mean value (<Bver>) is calculated and the standard deviation (SD) is found (see Table 1). At present, to draw the contour of the U-PU boundary on the intensity map, it is recommended to choose a level at the boundary of 50% of the intensity of the surrounding calm Sun surface. However, using the method of minimizing the distance between the contours Ic and Bver ([8]; Fig. 4), the authors obtained the optimal value of 53%, (Ic = 0.53 Iqs) and the corresponding value Bver = 1639G, which is almost 250G lower than the previously declared value in 1867G (see [3]). A deviation of 3% in the construction of the Ic-contour, which at first glance is insignificant, leads to strong changes in the values of Bver, due to the large field gradients near the U-PU boundary. To independently determine the level of this boundary, we used one of the mathematical methods for highlighting an object against the surrounding background - method to the Father [9]. When dividing the pixel field into classes, the clusterization problem, ie. the choice of an object as such, in general, is reduced to maximizing the interclass variance. Having received a set of objects, we select from them the ones we need for research: spots, pores, etc.

**3. Results**
Twelve "correct" single sunspots were investigated in the range of magnetic field values from 2 to 3 kG. In fig. 1 shows the Otsu Ic contour and 50% threshold (left) as an example for spot 1256/11388, as well as the Ic contour transferred to the magnetogram (right). It turned out that the outer boundary of the sunspot shadow is systematically higher than 50% of the level of the surrounding background and the excess of this threshold is different in each specific case. In fig.

2 shows a typical change in Bver along the contour in the magnetogram. Average field = 1628 G, but its value is not constant at all, it has a scatter of values up to 300 G (SD = 108).
A general summary of the research results is presented in Table 1.

**Table 1.**
Otsu threshold, mean vertical field on the Ic-contour and its deviation.

| N | SHARP/ NOAA | Data | Time CM (UT) | Otsu Contour Flux | Background <Flux> | R, % | <IBverI> Contour, G | SD, G | IB$_r^{max}$I, G |
|---|---|---|---|---|---|---|---|---|---|
| 1 | 1256/11388 | 2012/01/02 | 06:24 | 29700 | 56850 | 0.52 | 1628 | 102 | 2625 |
| 2 | 1256/11389 | 2012/01/03 | 08:00 | 30100 | 57850 | 0.52 | 1520 | 65 | 2483 |
| 3 | 1278/11391 | 2012/01/08 | 20:00 | 31035 | 59440 | 0.52 | 1480 | 58 | 2654 |
| 4 | 1321/11401 | 2012/01/20 | 17:00 | 26340 | 54880 | 0.48 | 1582 | 80 | 2941 |
| 5 | 1321/11402 | 2012/01/21 | 06:00 | 30530 | 57800 | 0.53 | 1575 | 115 | 2570 |
| 6 | 1350/11413 | 2012/02/01 | 15:00 | 27670 | 57500 | 0.48 | 1680 | 68 | 2645 |
| 7 | 1399/11420 | 2012/02/18 | 18:00 | 29755 | 58400 | 0.51 | 1520 | 77 | 2490 |
| 8 | 1422/11423 | 2012/02/29 | 07:00 | 29728 | 57100 | 0.52 | 1555 | 85 | 2579 |
| 9 | 1603/11466 | 2012/04/24 | 17:45 | 29825 | 58420 | 0.51 | 1420 | 68 | 2023 |
| 10 | 1644/11477 | 2012/05/14 | 18:00 | 29680 | 57534 | 0.52 | 1629 | 108 | 2720 |
| 11 | 1653/11479 | 2012/05/17 | 00:00 | 30068 | 58225 | 0.52 | 1690 | 86 | 2650 |
| 12 | 1677/11486 | 2012/05/24 | 00:00 | 29276 | 57725 | 0.51 | 1635 | 72 | 1962 |

In the columns of the table in order: the number of the patch registration by SHARP and NOAA, the date and time of the passage of the object (spot) through the central meridian (CM), the intensity value on the Ic-contour, calculated by the Otsu method [8], averaged over 4 regions background intensity near the spot, flux ratio (R), mean magnetic field value on the contour and its standard deviation (SD). As can be seen from the table, the field value <Bver> averaged over the Ic contour varies slightly (1.5–1.7 kGs), while the vertical component of the magnetic field itself, in a particular spot, changes significantly along the determined contour (the value of dis - Persia is quite high). These variations in the magnetic field at the U-PU boundary appear to directly reflect the well-known fine filamentous structure of the sunspot penumbra magnetic field. The border of the shadow of the spot, with the exception of some spots, is systematically higher than 50% of the level of the surrounding background and must be determined on a case-by-case basis. The average value R = 0.52 we found is close to the value determined in [9]. This leads to lower values than the value <Bver> = 1867G obtained in [1–3].

**4. Conclusions**
Using the mathematical method Otsu [8], it was shown that in a wide range of spot fields (from 2 kG to 3 kG), the average value of the field <Bver> along the contour changes weakly (1.5–1.7 kG), while this vertical component of the magnetic field noticeably changes along the determined Ic-contour (SD is high enough) due to the presence of a fine filamentous structure of the magnetic field of the sunspot penumbra. The U-PU boundary turns out to be systematically above 50% of the level of the ambient background and must be found for each specific case. This gives lower <Bver> values than the value of ~ 1867G given in [1–3]. For our sample of sunspots, we get <Bver> = 1576 +/- 82G.
It should be emphasized that we considered only fairly regular, well-developed sunspots that are at the stage of stable existence, with magnetic fields, which, according to the model of a shallow sunspot [10], provide the greatest stability margin for the magnetic system as a single whole. Therefore, there is no reason to extend the obtained conclusions to all sunspots in general and to consider the value of <Bver> at the "shadow-penumbra" boundary as some kind of solar physical constant.

**Referenses:**

1. Jurčák, J. // A&A, 2011, 531, A118.
2. Jurčák, J., Bello González, N., Schlichenmaier, R., Rezaei, R. // A&A, 2017, 597, A60.
3. Jurčák, J., Rezaei, R., González, N., Bello Schlichenmaier, R., Vomlel, J. // A&A, 2018, 611, L4.
4. Bobra, M. G., Sun, X., et al. (8 authors) // arXiv: 1404.1879v1 [astro-ph.SR], 7 Apr 2014.
5. Schou, J., Scherrer, P. H., Bush, R. I., et al. (total 21 authors) // Solar Phys., 2012, 275, 229.
6. Scherrer, P. H., Schou, J., Bush, R. I. (total 14 authors) // Solar Phys., 2012, 275, 207.
7. Calabretta, Mark R & Greisen, Eric W. arXiv: astro-ph / 0207413v1 19 Jul 2002.
8. Otsu, N.A. IEEE Transactions on Systems // Man and Cybernetics. 1979. V. 9. R. 62–66.
9. Schmassmann, M., Schlichenmaier, R. & González, N.B. arXiv: 1810.09358v2 [astro-ph. SR] 12 Nov 2018.
10. Soloviev A.A. & Kirichek E.A. // Astrophysics and Space Science, 2014, Vol. 352, No.1, 23-42.